\documentclass{webofc}

\usepackage{lineno,blindtext}
\usepackage{caption}
\usepackage{graphicx}
\usepackage{amssymb}
\usepackage{epstopdf}
\usepackage{amsmath,amsfonts,amsthm,amssymb}
\usepackage{bm}
\usepackage{subfig}
\usepackage{lipsum}
\usepackage{setspace}

\newcounter{daggerfootnote}
\newcommand*{\daggerfootnote}[1]{%
    \setcounter{daggerfootnote}{\value{footnote}}%
    \renewcommand*{\thefootnote}{\fnsymbol{footnote}}%
    \footnote[2]{#1}%
    \setcounter{footnote}{\value{daggerfootnote}}%
    \renewcommand*{\thefootnote}{\arabic{footnote}}%
    }

\begin{document}

\title{TeV-PeV neutrino-nucleon cross section measurement with 5 years of IceCube data}

\author{\firstname{Yiqian}
  \lastname{Xu}\inst{1}\fnsep\thanks{\email{yiqian.xu@stonybrook.edu}} on behalf of the IceCube Collaboration\protect\daggerfootnote{ http://icecube.wisc.edu/collaboration/authors/2018/04}
}

\institute{Department of Physics and Astronomy, Stony Brook University, Stony Brook, NY 11794-3800}

\abstract{

We present a novel analysis method for the determination of the neutrino-nucleon Deep Inelastic Scattering (DIS) cross section in the TeV - PeV energy range utilizing neutrino absorption by the Earth.
We analyze five years of data collected with the complete IceCube detector from May 2011 to May 2016. This analysis focuses on electromagnetic and hadronic showers (cascades) mainly induced by electron and tau neutrinos. The applied event selection features high background rejection (< 10\% background contamination below 60 TeV, background free above 60 TeV) of atmospheric muons and high signal efficiency (\char`\~ 80\%). The final neutrino sample consists of 4808 events, with 402 events above 10 TeV reconstructed energy. An unfolding method was applied to enable the mapping from reconstructed cascade parameters such as energy and zenith to true neutrino variables. The analysis was performed assuming isotropic astrophysical neutrino flux, in seven energy bins, and in two zenith bins ("down-going" from the south-hemisphere and "up-going" from the north-hemisphere). The ratio of down-going to up-going events (which are absorbed by the Earth at high energies) is sensitive to the neutrino-nucleon cross section but insensitive to the astrophysical neutrino flux uncertainties.

The neutrino-nucleon DIS cross section thus inferred is consistent with the Standard Model expectation within the uncertainties.

}

\maketitle

\noindent\textbf{Introduction}\hspace{.1\textwidth} 
IceCube is a one cubic kilometer size neutrino detector located at the geographic South Pole \cite{icecube_2017} . It is capable of detecting all-sky neutrinos of all flavors from GeV to EeV energies \cite{IceCube_instrum}. In 2012, it discovered the flux of extraterrestrial neutrinos in the TeV - PeV energy range \cite{HESE_2013}, alongside measuring neutrinos of known atmospheric origin. Direct neutrino-nucleon cross section measurements with fixed target experiments cover the neutrino energy range up to 370 GeV, while IceCube enables cross section measurements well beyond this energy. In 2017 IceCube published the first measurement of the neutrino-nucleon cross section (using $\nu_{\mu}$ coming from the Northern Sky) at neutrino energies above 6 TeV  \cite{sandy_xsec}. In this analysis we will present a novel analysis method \cite{D.Hooper_PRD} and the preliminary result of a cross section measurement using neutrino induced cascades from all-sky collected by IceCube from May 2011 to May 2016.  

\vspace{0.2cm}
\noindent\textbf{Event Selection}\hspace{.1\textwidth} Events observed in IceCube can be classified into two groups: Cascade and Track by their topology. Cascade events are induced by $\nu_{e}, \nu_{\tau}$ charged current interactions and all flavor neutrino neutral current interactions. The emitted light pattern for cascade events are shower-like. Track like events are induced by $\nu_{\mu}$ charged current interactions. The signal for the present analysis are cascade events with interaction vertices within the detection volume of IceCube, and the background are muons from cosmic rays. The majority of the events seen at trigger level are background events. To distinguish them from signal events we studied their characteristic variables distribution using Monte Carlo simulations. The codes \textit{MuonGun} \cite{muongun} and \textit{neutrino-generator} \cite{nugen} are used respectively to generate muons in ice with the cosmic ray composition from \cite{gaisser_h3a} and neutrinos. The atmospheric neutrino flux is modeled with HKKMS06 \cite{conv} (for the conventional part which is produced by charged pion and kaon decay) and BERSS \cite{prompt} (for the prompt part which is produced by heavier meson decay), with "self-veto" effect \cite{self_veto} taken into consideration. The normalization: $(1.5^{+0.23}_{-0.22})\times10^{-18}\mathrm{\ GeV^{-1}s^{-1}sr^{-1}cm^{-2}}$, and spectral index: -2.48$\pm$0.08 of an unbroken single power law ($\Phi_{astro}=\Phi_{0}\times(E_{\nu}$/100 TeV$)^{-\gamma})$ fit to the astrophysical neutrino flux is the preliminary fit result from the IceCube 4-Year Cascade analysis \cite{cscd_ICRC2017}\cite{hans_thesis}. 

\begin{figure}[h!]
\centering
\subfloat{\includegraphics[scale=0.15]{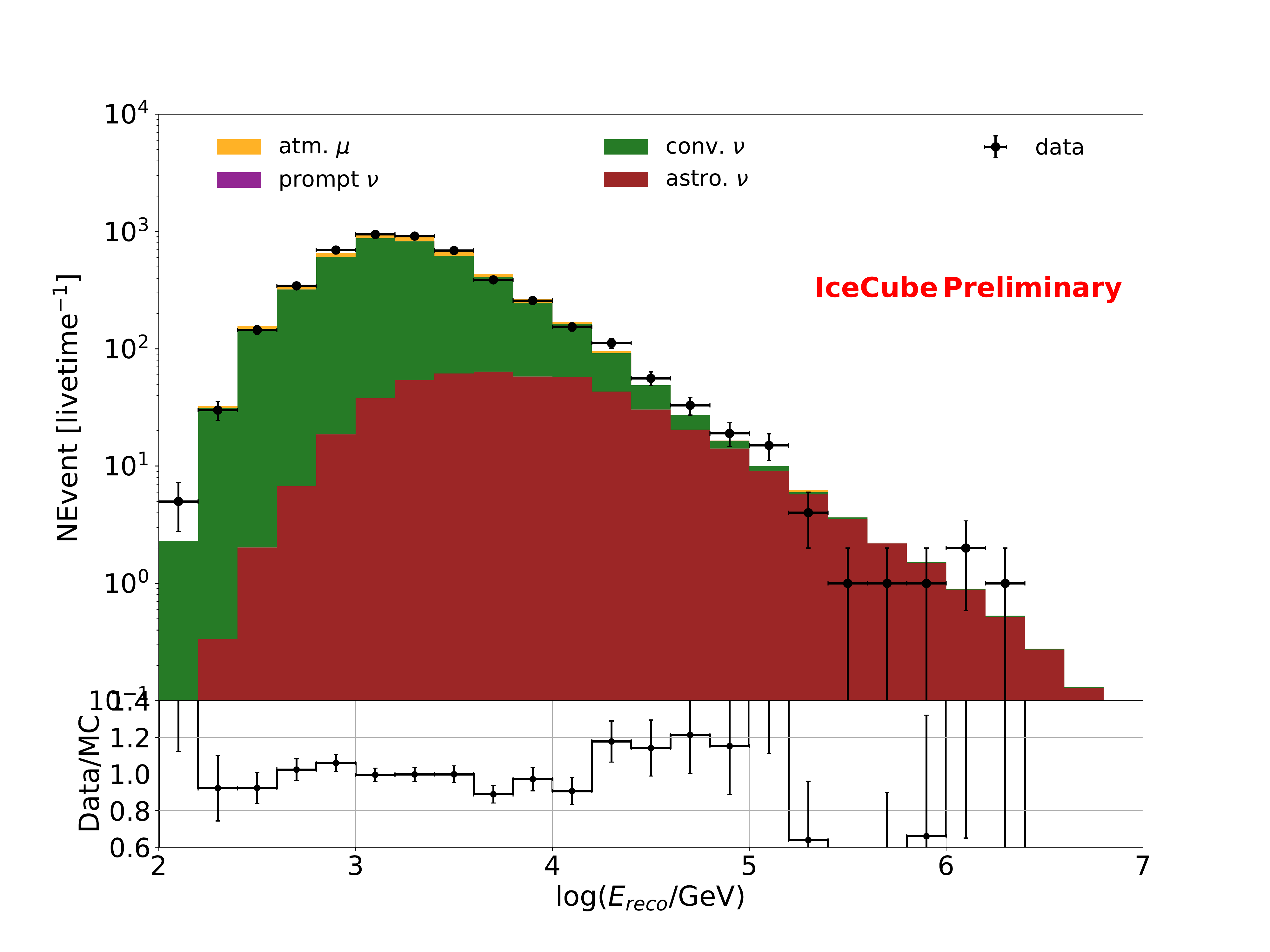}}
\subfloat{\includegraphics[scale=0.15]{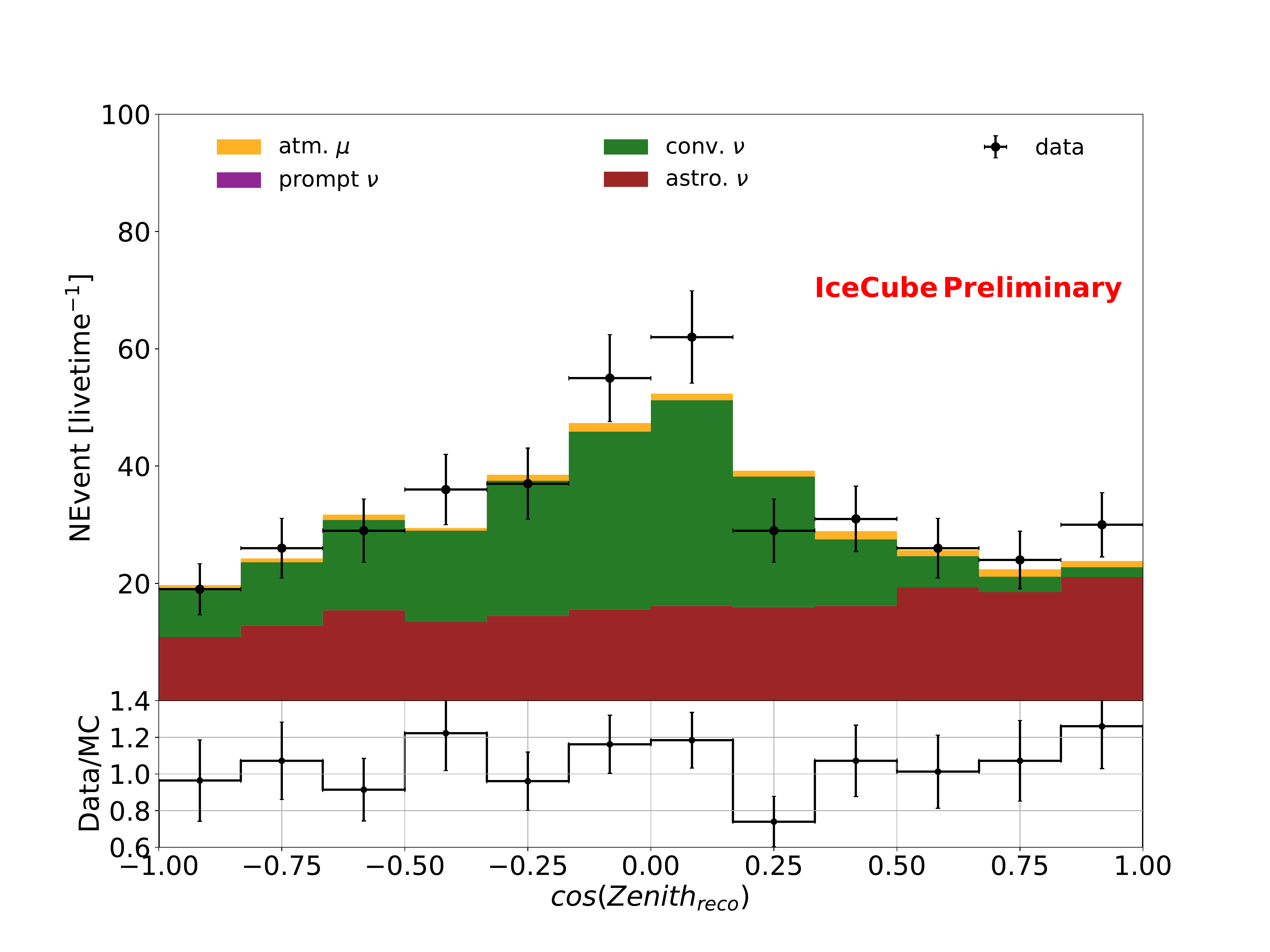}}
\caption{Left: reconstructed energy distribution for data and Monte Carlo. Right: reconstructed zenith distribution for data and Monte Carlo (above 10 TeV) .}
\label{fig:variable_dist}
\end{figure}

The event selection is conducted in two energy ranges: the high energy part ($E_{reco}$ > 60 TeV) where a 2D straight cut strategy is applied and the low energy part ($E_{reco}$ <= 60 TeV ) where a machine learning algorithm called Boosted Decision Tree is applied. After the selection, there are 4808 events left. The study done on Monte Carlo simulations shows that the high energy part is background free, while the background contamination in the low energy part is lower than 10\%, and the signal efficiency is 80\%. Fig.\ref{fig:variable_dist} shows the reconstructed energy and zenith distribution for experimental data and Monte Carlo simulations at the final level. 

\begin{figure}[h!]
\centering
\subfloat{\includegraphics[scale=0.35]{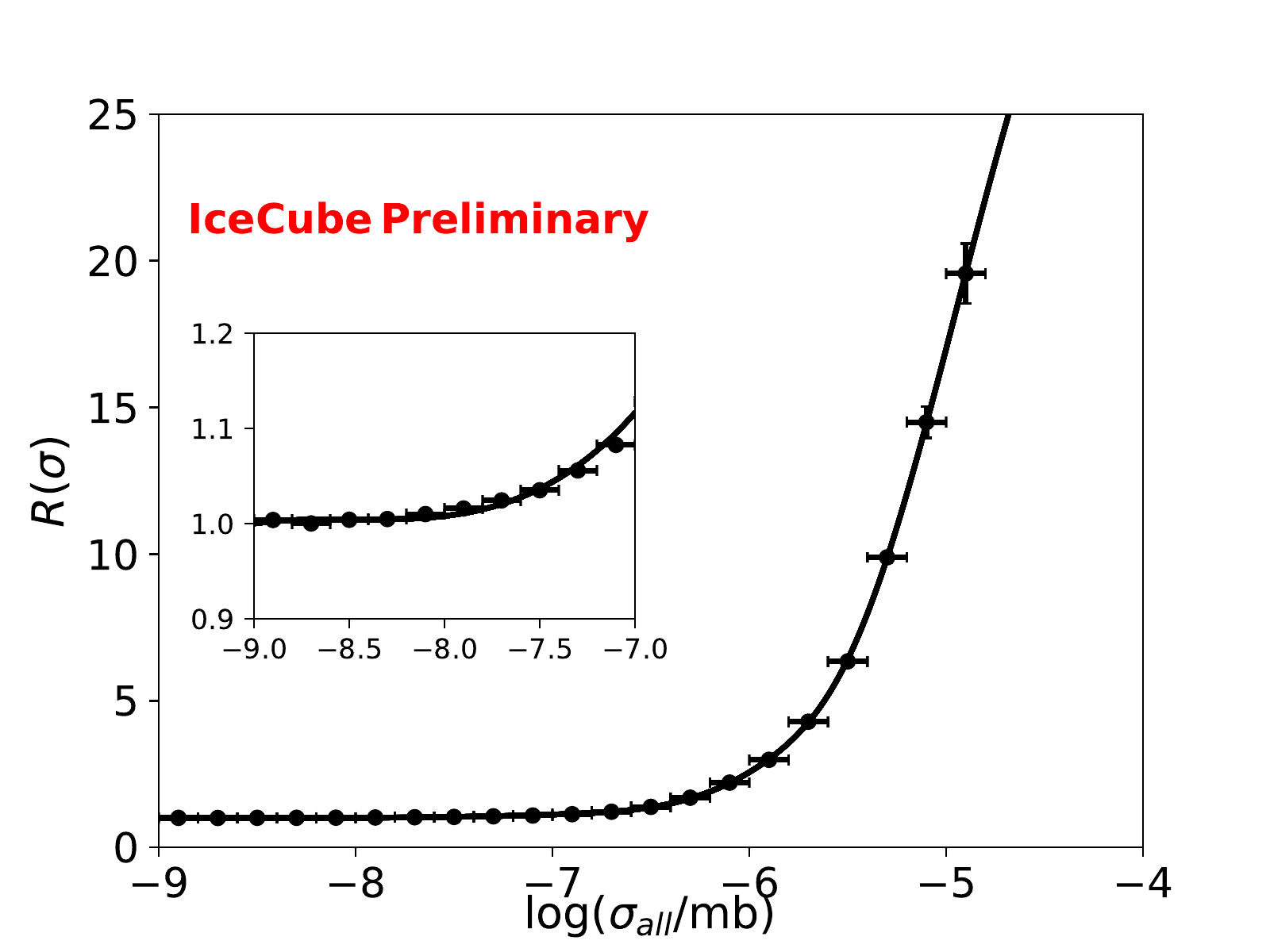}}
\caption{Ratio vs. cross section of down-going and up-going events. (generated by Monte Carlo)}
\label{fig:xsec_vs_ratio}
\end{figure}

\vspace{0.2cm}
\noindent\textbf{Analysis Method and Uncertainty Estimation}\hspace{.1\textwidth}  This analysis is inspired by the theoretical approach of Ref. \cite{D.Hooper_PRD}. While IceCube detects neutrinos from the Southern Sky without attenuation, a fraction of the neutrinos from the Northern Sky traveling through the Earth get absorbed by the Earth before they can reach the detector \cite{xsec_ICRC2017}. This fraction depends on the cross section \cite{D.Hooper_PRD} and can be quantified by the ratio of down-going (Southern Sky) to up-going events (Northern Sky). The number of events observed at the detector is correlated with neutrino flux and neutrino interaction cross section. We assume an isotropic astrophysical neutrino flux and we know that the atmospheric neutrino flux is symmetric around the horizon \cite{conv}, \cite{prompt}. The ratio of down-going to up-going events is thus determined by the cross section only. The relationship between this ratio and the cross section can be calculated using Monte Carlo simulations and is shown in Fig. \ref{fig:xsec_vs_ratio}. The Earth density model used in the simulation is PREM \cite{PREM_1981}.

Experimental data are assigned in different reconstructed energy bins and separated into down-going and up-going events within each bin to calculate the ratio. We reconstruct the deposited amount of energy. In $\nu_{e}$ and $\nu_{\tau}$ charged current interaction, the deposited energy is close to the original neutrino energy, however in neutral current interactions and $\nu_{\mu}$ charged current interactions, part of the energy is carried by the out-going particle, therefore the the deposited energy is smaller than the neutrino energy. In addition there are resolution effects:  $\sigma_{E}$ \char`\~15\%, $\sigma_{\theta}$ \char`\~$10^{o}-15^{o}$ \cite{energy_reso}, hence the reconstructed zenith distribution is not the true distribution. To account for these effects, a 2D iterative unfolding method \cite{unfolding} is applied in both energy and zenith space. Since the energy deposition from interactions is flavor dependent, we assume a 1:1:1 flavor ratio for the astrophysical flux in this analysis. 

Iterative unfolding is based on Bayes' Theorem. The probability of having reconstructed energy $R_{j}$ given neutrino energy/zenith $T_{i}$ ($P(R_{j} | T_{i})$) is known, and the probability of having neutrino energy/zenith $T_{i}$ given reconstructed energy/zenith $R_{j}$ ($P(T_{i} | R_{j})$) is of interest. The equations that connect the two probabilities are:
 
\begin{align}
P(T_{i}|R_{j})^{(k)}=\frac{P(R_j{|}T_{i})N(T_{i})^{(k)}}{\sum_{i}P(R_{j}|T_{i})N(T_{i})^{(k)}},\\
N(T_{i})^{(k+1)}=\sum_{j} P(T_{i}|R_{j})^{(k)}N(R_{j})/\alpha_{j},
\end{align}
\setlength{\belowdisplayskip}{0pt} \setlength{\belowdisplayshortskip}{0pt}
\setlength{\abovedisplayskip}{0pt} \setlength{\abovedisplayshortskip}{0pt}
where $N(T_{i})^{(k+1)}$ is the distribution in neutrino energy/zenith space after k+1th iteration, $N(R_{j})$ is the distribution in reconstructed energy/zenith space, and $\alpha_{j}$ is a correction factor for acceptance losses \cite{unfolding}. $N(T_{i})^{(0)}$ is chosen to be Monte Carlo truth. 

Statistical uncertainty on the unfolded ratio is estimated using Markov Chain Monte Carlo (MCMC) to sample events in reconstructed space and unfold each sampled event to get the posterior distribution in neutrino space. The 68\% confident interval in the posterior distribution is taken as the statistical uncertainty. 

\begin{figure}[h!]
\centering
\subfloat{\includegraphics[scale=0.35]{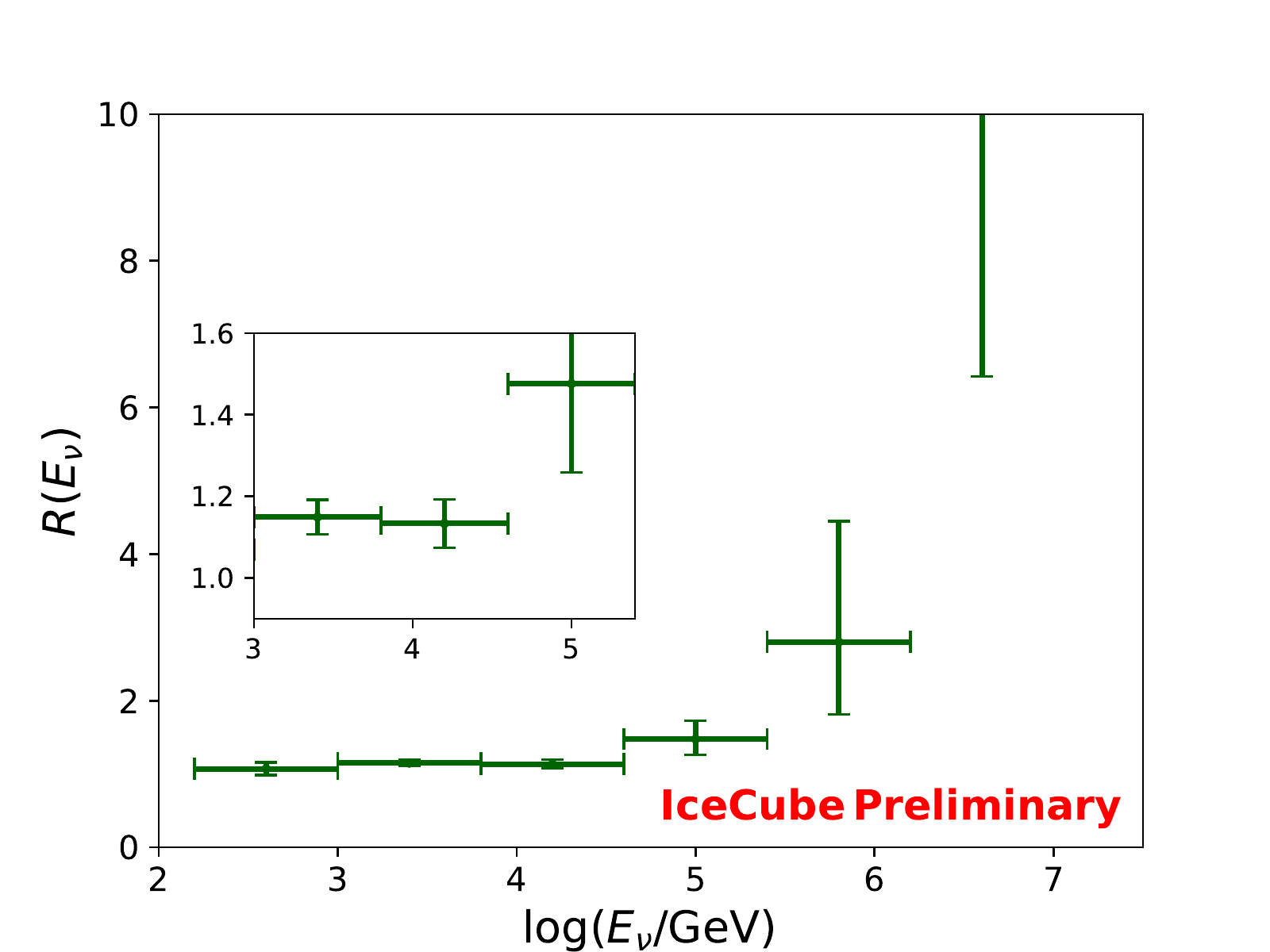}}
\subfloat{\includegraphics[scale=0.35]{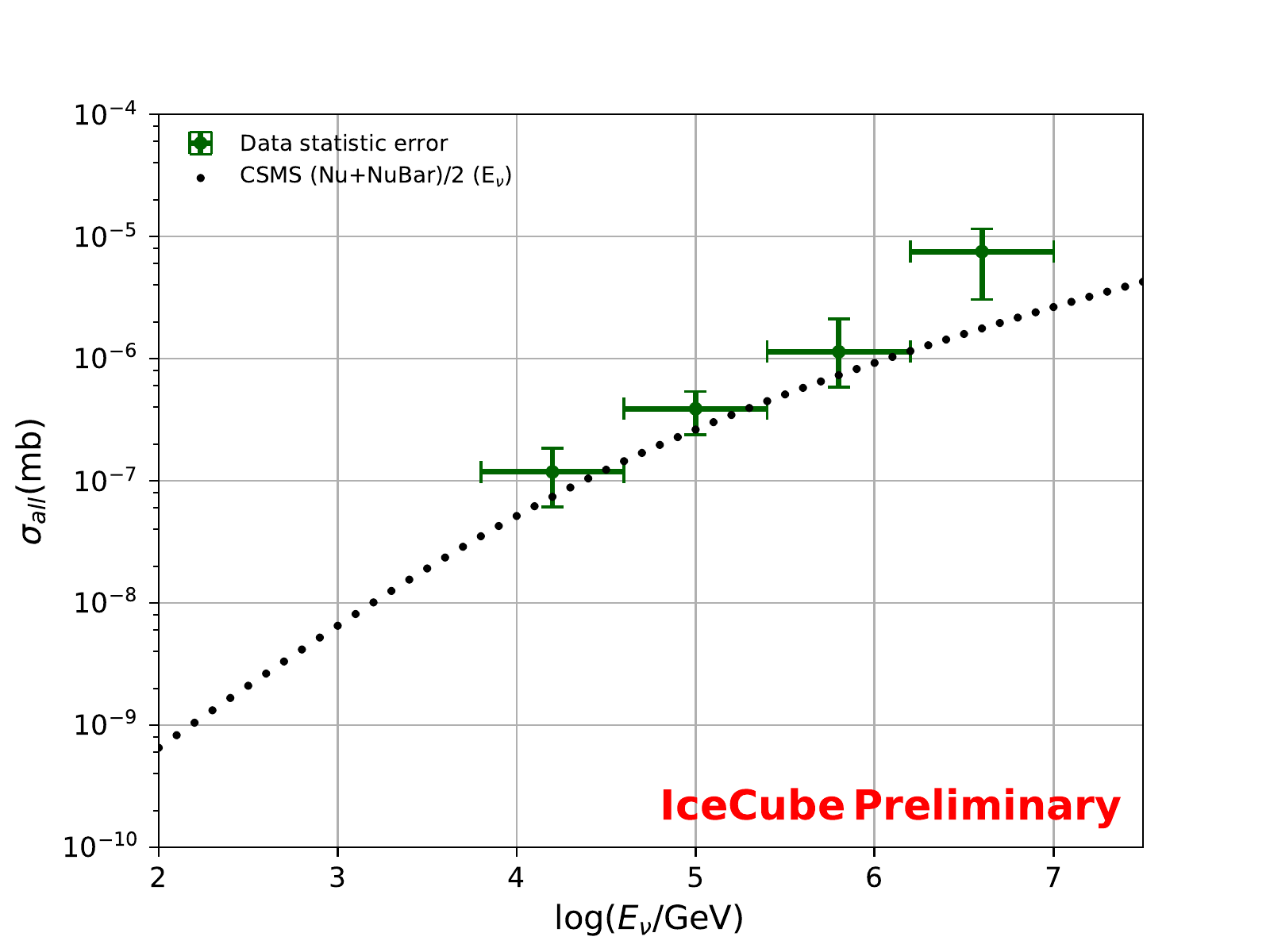}}
\caption{Left: the ratio of down-going to up-going events for 5 years of data vs. neutrino energy. Uncertainties are statistical only. Right: the inferred cross section vs. neutrino energy with statistical uncertainties only.  }
\label{fig:energy_vs_ratio}
\end{figure}

Due to the \textit{self-veto} \cite{self_veto} effect in atmospheric neutrinos flux and event selection, the detector has a different efficiency for detecting down-going and up-going events. Without correction, the ratio of down-going events to up-going events will depend not only on the Earth absorption, but also on the detector acceptance. A Monte Carlo study enables the separation of these effects, and a correction factor is applied after unfolding to calculate the ratio that reflects Earth absorption only. The ratio in neutrino energy bins with the statistical uncertainty estimated with the method stated above is shown in the left plot in Fig. \ref{fig:energy_vs_ratio}. For each neutrino energy bin, we take the ratio in the bin and find the corresponding cross section value using the ratio vs. cross section curve in Fig. \ref{fig:xsec_vs_ratio}. In this way, we obtain the neutrino cross section vs. energy with statistical uncertainty as shown in Fig.\ref{fig:energy_vs_ratio} right plot. The measured neutrino-nucleon cross section is the sum of neutral current and charge current interaction cross sections, averaged over neutrino and anti-neutrino since the two can not be distinguished in IceCube.

\begin{figure}[h]
\centering
\subfloat{\includegraphics[width=0.4\textwidth]{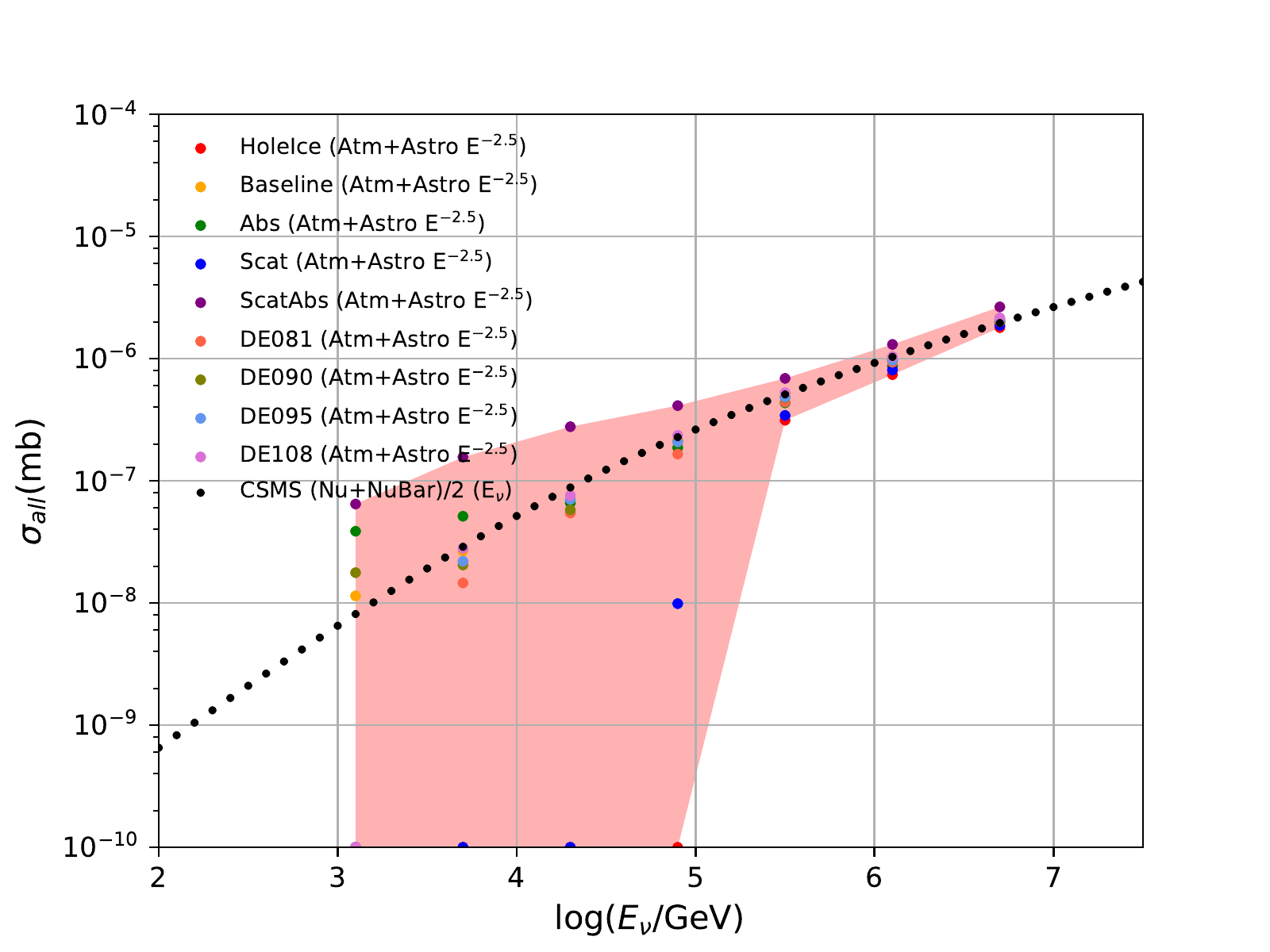}}
\subfloat{\includegraphics[width=0.4\textwidth]{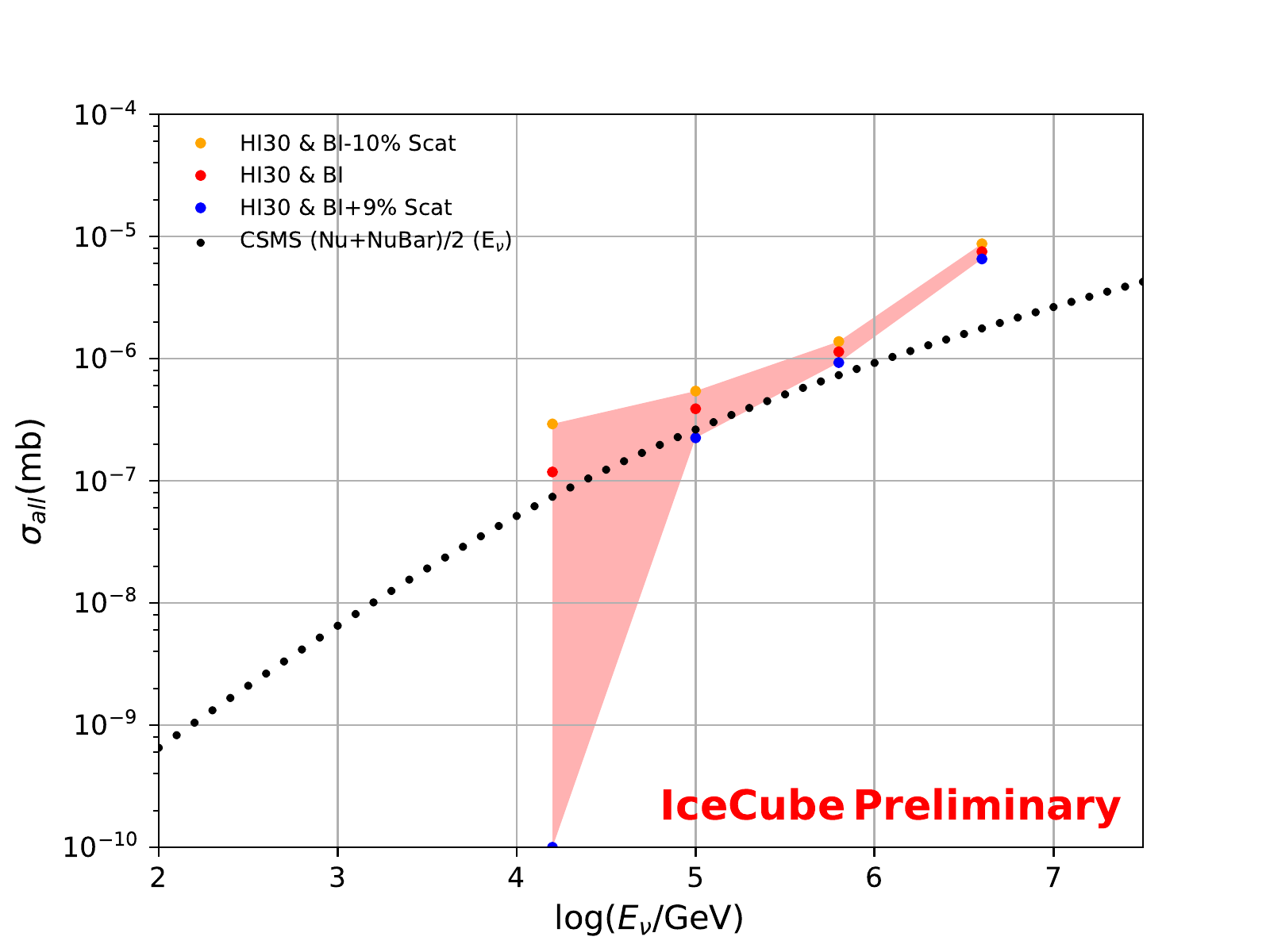}}
\caption{Left: Systematic effect study using Asimov data. (legend: HoleIce: scattering length 30cm, Abs: +10\% absorption, Scat: +10\% scattering, ScatAbs: -7\%scattering and -7\% absorption, DEXXX: DOM efficiency XXX\%)  Right: Systematic uncertainty of the cross section measurement using 5 years of IceCube data. (legend: HI30:scattering length 30cm, BI +/-X\% Scat: changing bulk ice scattering by +/-X\%)}
\label{fig:sys_error}
\end{figure}

The systematic effects considered in this analysis are associated with photon scattering and absorption in the South Pole ice, Digital Optical Module (DOM) efficiency, self-veto \cite{self_veto} effect, and the astrophysical neutrino flux. Due to the method of taking a ratio, the uncertainty associated with the astrophysical flux is negligible. The uncertainty on self-veto parameters is small enough that its contribution to the cross section measurement result is negligible as well. For the other systematic effects, Monte Carlo simulations have been generated to study their impact on the result. A study done on Asimov data shows that the most significant systematic effect is photon scattering in the ice (Fig.\ref{fig:sys_error} Left).  There are two parts of the South Pole ice that are of interest: bulk ice (the pristine glacial ice), hole ice (the refrozen ice around the DOMs after installation). Due to the lack of the hole ice scattering variation Monte Carlo simulation, we currently only consider the variation of the bulk ice scattering as systematic effect here. 

The systematic effects change the mapping between reconstructed and true variables. Data is unfolded with matrices built with the simulation of different systematic effects. The corresponding cross section measurement result is shown in Fig. \ref{fig:sys_error} Right. The red band covers the range of the preliminary systematic uncertainty of this measurement. 

\begin{figure}[h]
\centering
\subfloat{\includegraphics[scale = 0.5]{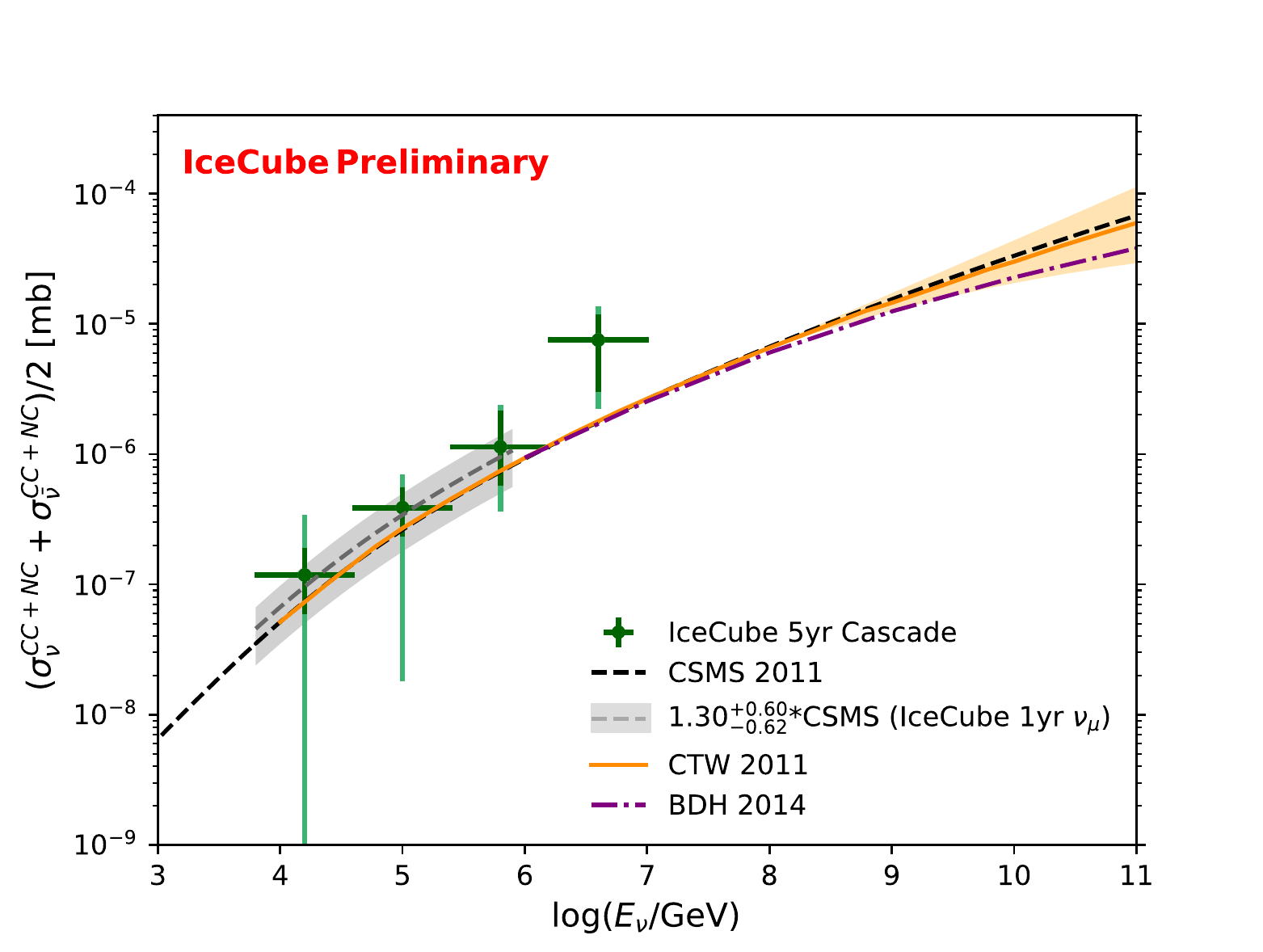}}
\caption{The energy dependence (in the 6 TeV-10 PeV range) of the neutrino-nucleon cross section measurement with statistical and preliminary systematic uncertainties in comparison with theoretical models}
\label{fig:energy_vs_xsec}
\end{figure}

\noindent\textbf{Result}\hspace{.1\textwidth} The cross section measurement with combined statistical uncertainty (Fig. \ref{fig:energy_vs_ratio} right) and preliminary systematic uncertainty (Fig.\ref{fig:sys_error} right) is shown in Fig.\ref{fig:energy_vs_xsec} in comparison with the previously published IceCube 1 year $\nu_{\mu}$ cross section measurement ($1.3^{+0.60}_{-0.62}\times \sigma_{SM}^{CSMS}$), and several theoretical DIS cross section calculations \cite{csms}, \cite{CTW_2011}, \cite{BDH_2014}. This result covers the neutrino energy range from 6 TeV to 10 PeV. It is consistent with all the Standard Model cross section calculations and the previously published IceCube result within the uncertainties. 

\vspace{0.2cm}
\setstretch{1.25}\noindent\textbf{Summary and Outlook}\hspace{.1\textwidth} We have developed a novel analysis method to measure neutrino-nucleon cross section in the 6TeV - 10PeV neutrino energy region using 5 years of IceCue cascade sample. A 2D iterative unfolding method has been used. Statistical uncertainty has been estimated using MCMC. Preliminary systematic uncertainty has been estimated as well. The result shows consistency with Standard Model cross section calculations within the uncertainties. With more years of data accumulation and the proposed upgrade of the detector, IceCube has the potential to probe in higher energy regime (>1 EeV) where there is sensitivity to new physics.

\end{document}